\documentclass[a4paper,longmktitle]{cas-sc}

\usepackage[numbers,sort&compress]{natbib}

\def\tsc#1{\csdef{#1}{\textsc{\lowercase{#1}}\xspace}}
\tsc{WGM}
\tsc{QE}
\tsc{EP}
\tsc{PMS}
\tsc{BEC}
\tsc{DE}

\begin{document}
\let\WriteBookmarks\relax
\def\floatpagepagefraction{1}
\def\textpagefraction{.001}
\shorttitle{Moderate death rates can be beneficial for the evolution of cooperation}
\shortauthors{EJS Junior et~al.}

\title [mode = title]{Moderate death rates can be beneficial for the evolution of cooperation}

\tnotetext[1]{This document is the results of the research
   project funded by CNPq and FAPEMIG}

\author[1]{Elton J. S. J\'unior}
\address[1]{Centro de Federal de Educa\c c\~ao Tecnol\'ogica de Minas Gerais, CEP 35790-000, Curvelo - MG, Brazil.}

\author[2]{Marco A. Amaral}
\address[2]{Instituto de Humanidades, Artes e Ci\^encias, Universidade Federal do Sul da Bahia, CEP 45638-000, Teixeira de Freitas - BA, Brazil.}

\author[3]{Lucas Wardil}
\cormark[1]
\ead{wardil@fisica.ufmg.br}
\address[3]{Departamento de F\'isica, Universidade Federal de Minas Gerais, Caixa Postal 702, CEP 30161-970, Belo Horizonte - MG, Brazil.}
\cortext[cor1]{Corresponding author}

\begin{abstract}
Spatial structure is one of the simplest and most studied ecological factors that affect the evolution of cooperation. It has been shown that spatial reciprocity promotes cooperation due to the formation of cooperative clusters, which provide mutual support against defectors. The usual assumption is that of constant population size, where no density-related effect is possible. Here, we extend the investigation of density-related effects on the evolution of cooperation.  We integrate evolutionary game theory to the Lattice Lotka-Volterra Model. In our model,  the birth rate depends on the local density and on the payoff accumulated in the interactions.  We characterize the evolution of cooperation in terms of the coexistence and the extinction of the types. The main result is that cooperation is most favoured at moderate levels of the death rate.  Interestingly,  defectors are extinct at values of the death rate that are lower than that at which cooperators are extinct. When death knocks the door,  defectors are the first to perish, whereas cooperators stand longer due to mutual support.

\end{abstract}

\begin{highlights}
\item Game theory is incorporated in the Lattice Lotka-Volterra model.
\item Density related effects are taken into account in the reproductive rate.
\item Cooperation level reaches its peak at intermediate values of the death rate.
\end{highlights}


\maketitle

\section{Introduction}

Evolutionary game theory has been used  to study frequency-dependent selection \cite{freq,bio-game}. The essence of the theory is to define a Darwinian dynamic that determines the evolution of individual types in populations where the fitness depends on the frequency of the types. The specifics of the biological process are often omitted to provide a clearer picture of the arms race, wherein strategies with high fitness propagate at higher rates \cite{nowak-book}. The simplest approach is to study the evolution in well-mixed populations, where the fitness is determined by the payoff accumulated over random interactions. The classical Darwinian definition of fitness -- the ratio between the prevalence of one type between two generations -- is consistent with the definition of fitness in the Wright-Fisher model \cite{chalub}, which has been widely used by the genetic population theorists. However, to attain analytical tractability, the  Moran process is often used \cite{nowak-book,chalub,genetic-book}. The downside of both models is that the population size is constant and no density-dependent effect is possible. Since these two models are defined using finite population, the  analysis typically focuses on the fixation probability of an invading mutant and on the fixation time of the mutant. In general,  to attain simple and general results, many ecological features are ignored  \cite{replicator-book,replicator2-book}.  Density-dependent effects can drastically change the outcome of the population evolution, and a deeper understanding of such mechanisms is important, as done by Wang et al in \cite{wangfreq, wangfreq2}, where it was show that the lattice percolation point is related to the optimal cooperation enhancement.

Cooperation is quite puzzling because cooperators incur reproductive costs to bestow benefits on others. However, we still find numerous forms of cooperative behavior in nature \cite{fehr}. Hence, there must be mechanisms providing indirect benefits to cooperators that support their existence. Many mechanisms have been studied and comprehensive reviews can be found in the literature \cite{five-rules,fletcher-doebeli,volunteer,reviewperc, ciclic} . One ecological factor that provides indirect benefits to cooperators is viscous population structure: individuals, and their descendants, are spatially structured and have a low rate of dispersion \cite{viscous1}. If the offspring does not disperse, the cooperators can  out-compete the defectors due to the formation of islands of cooperation where mutual support is provided. Interestingly, if the viscosity is weakened and there are vacant sites, so the individuals are allowed to move, the cooperators can invade a population of defectors \cite{mobility}.

In a revamp of the viscous population mechanism,  Martin Nowak elaborated on what was later called network, or spatial, reciprocity \cite{spatial,szabo}. Network reciprocity appears in populations where individuals, which are represented by nodes of a graph, interact with fixed neighbours, determined by the links of the graph. Since strategies spread only within the neighborhood, clusters of cooperators emerge, making cooperators strong enough to beat defectors and recover from eventual invasions by defectors \cite{cluster}. To keep simplicity, most studies focus on  birth and death processes where new strategies appear only in sites that are freed by a death event \cite{simple-rule}, keeping the population size constant and ignoring population density variation effects. 

The coupling between birth and death events in mathematical models, although allowing analytical tractability, is too restrictive. Demographic stochasticity is  easily modelled if the population size is not fixed and death and birth events are not coupled. In this way, the fluctuation of the average fitness  reflects in the fluctuation of the population size. Population size fluctuations account not only for the possibility of extinction, but also for non-trivial phenomena. For example, in small populations, cooperators are less prone to go extinct than defectors \cite{demo}.  Additionally, if spatial structure is considered, the variation of population size creates density-related effects. For example,  increasing population density has an effect on reproductive rates that, together with spatial diffusion, promotes coexistence of cooperators and defectors in static or dynamic patterns, including spatial chaos of ever-changing configurations \cite{diffusion}.

The  Lattice Lotka-Volterra Model (LLVM) is an extension of the traditional Lotka-Volterra model where birth and deaths are naturally decoupled  \cite{LLVM}. The inclusion of spatial structure in the Lotka-Volterra model  opens the possibility  to study population size fluctuations and density-related effects  on the evolution of cooperation. In the original LLVM model, the cooperative behavior affects the death rate and an invasion analysis has been carried out, showing how invasion depends on the local density of cooperators \cite{LLVM}.  Other models have been developed to study the general behavior of predator-prey systems in square-lattice, where phase transitions and oscillatory behavior have been observed \cite{antal2001}. Here, we elaborate on  the LLVM to study the evolution of cooperation within the usual framework of evolutionary game theory \cite{szabo}. 

In our model, the population is structured on a square lattice, with  individuals playing the prisoner's dilemma game  with its   first neighbors (von Neumann neighborhood \cite{nowak-book}). The birth rate is determined by the combination of the accumulated payoff over the interactions with nearest neighbours and the local population density. Death is a random event determined by a constant rate. In a short time interval, an individual may reproduce, die or remain as it is. Thus, population size can vary and empty sites can exist. 
We found that the variation of parameters can lead to the extinction, coexistence or fixation of both cooperators and defectors. In a certain region of the parameter space characterized by  coexistence, we found for finite systems the typical oscillatory behaviour of Lotka-Volterra models. The most interesting is that the change of survival of cooperators and defectors is not trivial: the fraction of defectors decreases if the death rate increases, but the fraction of cooperators exhibits a maximum at moderate death rates. This result suggests that, if there is competition between cooperators and defectors, moderate death rates can be beneficial for cooperators. 

In the next section, we present our model. In the results section, we derive a mean-field approximation, which can be interpreted as the evolution in a well-mixed population. Then, we present and discuss the results of the simulation of populations structured on a square lattice. 

\section{The model}
\label{model}

The population is structured on a square lattice with periodic boundary conditions. Each individual occupies one site of the lattice and interacts with his four nearest neighbors,  that is, we consider the von Newmann neighborhood. Individuals can adopt cooperation (C) or defection (D), interacting in a Prisoner's Dilemma game. Both players receive $R$ (reward) for mutual cooperation or $P$ (punishment) for mutual defection. If they chose different strategies, the cooperator receives $S$ and the defector receives $T$ (temptation). To simplify the analysis, we study the weak version of the prisoner's dilemma, defined by $T>1$, $R=1$, and $P=S=0$.

The evolutionary process is driven by birth and death events. The death rate is constant and uniform: all individuals have the same chance to die. The birth rate is determined by two components. First, the payoff accumulated in the interactions with the four nearest neighbours, plus a constant basal birth rate. Second, a density-related effect that decreases the birth rate if the local density is high. The offspring occupies an empty site in the neighbourhood of the predecessor.  
The birth and death process is implemented as follows. An individual $i$ is randomly chosen. Individual $i$ reproduces with probability $p_i^+$, dies with probability  $p_i^-$, or nothing happens with probability  $1-p_i^+-p_i^-$. The probabilities are defined by
\begin{equation}
p_i^+=\frac{w_i\left(b+U_i\right)}{b+U_{max}+d}
\end{equation}
and
\begin{equation}
p_i^-=\frac{d}{b+U_{max}+d},
\end{equation}
where $w_i$ is the fraction of empty sites in the neighbourhood of individual $i$, $U_i$ is the accumulated payoff in the interaction with nearest neighbours, $b$ is the basal birth rate, $d$ is the death rate, and $U_{max}$ is a normalization factor  (we set $U_{max}=4T$). If the individual $i$ reproduces, an offspring is placed on an empty site in the neighbourhood of $i$. If individual $i$ dies, the site becomes empty.  Therefore, empty sites can be created and the size of the population can change. The level of cooperation in the population is quantified by the fraction of sites occupied by cooperators, $u$; the fraction of sites occupied by defectors, $v$; and the fraction of empty sites, w. By definition, $u+v+w=1$.

\section{Results and discussions}
\label{results}

\subsection{Mean-field approximation}
The mean-field dynamical equations that describe the evolution of the fraction of cooperators, defectors  and empty sites in well-mixed populations are given by
\begin{eqnarray}
 \dot u &=& \left[w\left(b+U_C\right)-d\right]u, \\
 \dot v &=& \left[w\left(b+U_D\right)-d\right]v,
\end{eqnarray}
where $U_C=4u$ and $U_D=4Tu$ are the mean-field approximation of the payoffs. 

The fixed points satisfy $\dot u=0$ and $\dot v=0$. The four fixed points are summarized in Table \ref{mf}.
The Jacobian matrix, $J_{ij}=\frac{\partial \dot{x}_i}{\partial x_j}$, is given by
\begin{equation}
J=
\begin{bmatrix}
   (w-u)(b+4u)-d+4uw & -u(b+4u)  \\
   v(4T(w-u)-b) & (w-u)(b+4Tu)-d 
\end{bmatrix}.
\end{equation}
Depending on the eigenvalues of the Jacobian, the linear stability analysis can be used to classify the stability of the fixed points.

The first fixed point is the trivial solution  ${u=v=0}$ and ${w=1}$, which is simply the result of the extinction of the population. The Jacobian evaluated at this fixed point yields the eigenvalue $\lambda_1=b-d$ with multiplicity two. The point is stable if $b<d$ and unstable if $b>d$.
The second fixed point satisfies  $u=0$, $v\neq0$, and $w\left(b+U_D\right)-d=0$. In this case, we must have $d<b$. Since one of the eigenvalues of the Jacobian is zero, the linear stability analysis is inconclusive. However, along the line $u=0$ it is easy to check that the fixed point is stable.  That is, if only defectors are present, the system is stable at  the fixed point.
The third fixed point satisfies $v=0$, $u\neq0$, and $w\left(b+U_C\right)-d=0$. The analysis shows that the fixed point can be a saddle point or an unstable point  depending of the values of $T$ and $d$.  However, along the line $v=0$, the solution is always stable. That is, if only cooperators are present, the system is stable at  the fixed point.
 The fourth fixed point satisfies $u\neq0$, $v\neq0$, but must satisfy $T=1$. This solution represents coexistence of the two types  of players and vacancies along a  parametrized curve. Since one of the eigenvalues is zero, the linear stability analysis is inconclusive. However, a quick inspection on the vector field $(\dot{u},\dot{v})$ shows that the coexistence line is an attractor if the death rate is not too high.  
\begin{table}[pos=h]
\caption{Solutions of the mean-field approximation}
\label{mf}
\begin{tabular}{|c|c|c|c}
 \hline
 &Solution & Restriction \\ \hline 
 $P_1$&$u=0$, $v=0$ and $w=1$ & - \\ \hline
$ P_2$&$u=0$, $v=1-\frac{d}{b}$, $w=\frac{d}{b}$ & $d<b$ \\ \hline
 $P_3$&$v=0$, $w=\frac{b+4-\sqrt{\left(b+4\right)^2-4d}}{2}$, &  \\
 &$u=1-w$ & $0<d<b+3$ \\ \hline
 $P_4$&$u=u^*$, $v=1-w-u^*$, $w=\frac{d}{b+4u^*}$,  & $T=1$ \\
 \hline
\end{tabular} 
\end{table}

The mean field approximation can be interpreted as an approximation, where each individual interacts, on average, with four other random sites. It is important to note that in models with empty sites, or any type of density dependent effect, the mean field approximation has to explicitly consider such empty sites, e.g. utilizing the fraction of empty sites, $w$, as an active variable in the model. The state where the population is extinct is stable only if the death rate is greater than the birth rate, $b<d$; otherwise the population can escape extinction. The coexistence is possible only if defectors have no payoff advantage, that is, if $T=1$. In summary, as long as the other type does not invade,  an homogeneous population (of cooperators or of defectors) will be stable. Notice that the condition for the existence of an homogeneous population of defectors (fixed point $P_2$) is more stringent because defectors do not generate payoff.

\subsection{Simulations on a square-lattice}

We simulated the model in a square lattice of linear size $L=100$, which means $N=10000$ sites. The square lattice is initially fully occupied, with $50\%$ of cooperators and $50\%$ of defectors. The averages are taken on the stationary regime over 100 samples of different initial conditions. The baseline birth rate is given by  $b=1$ so that the only relevant parameters are the temptation to defect, $T$, and the death rate, $d$.  

\begin{figure}
\centering
\includegraphics[scale=0.7]{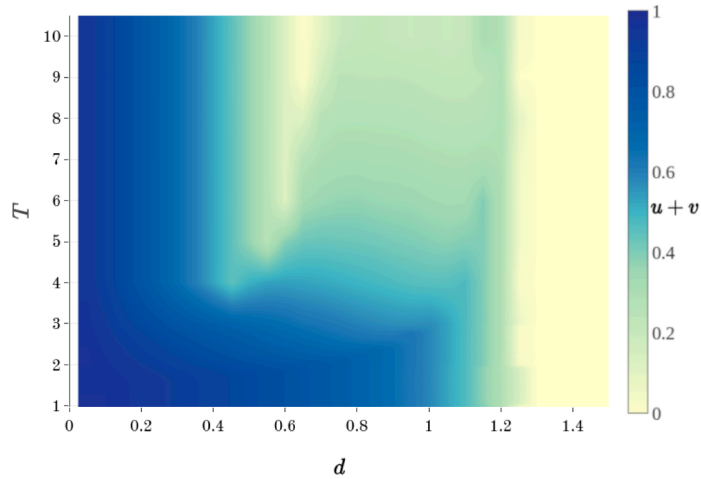}
\caption{Heat mapping encoding the population density in the square lattice for the whole $T-d$ plane. For high death rates, the population is extinct. The dependence of the population density on $T$ is better visualized in Fig. \ref{fig2}}
 \label{fig1}
\end{figure}

\begin{figure}
\centering
\includegraphics[scale=0.35]{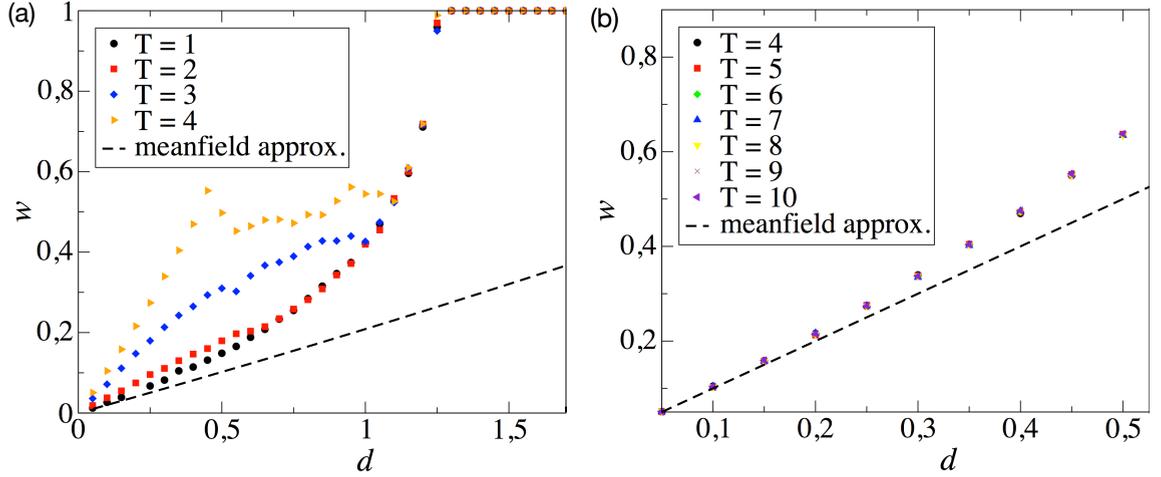}
\caption{Fraction of vacancies in equilibrium in the square lattice. The panel (a) shows the results for low values of $T$, where the population density is sensitive to the values of $T$. The simulations in panel (a) are compared to the mean field equilibrium where $u\neq0$ and $v=0$ (dashed lines). Panel (b) shows the results for higher $T$ values, where the population in not sensitive to the values of $T$. The mean field approximation in panel (b) corresponds to the equilibrium where $v\neq0$ and $u=0$. Error bars are comparable to the size of symbols.}
 \label{fig2}
\end{figure}

Figure \ref{fig1} shows that the population density decreases as $T$ and $d$ increases, with a weak non-monotonic dependence on $d$. If $d$ is sufficiently high ($d^*\approx1.2$), population is extinct. Surely, one would expect extinction for sufficiently high death rates. But notice that, due to the payoff produced by cooperators,  the population persists in the square lattice for $d>b$ (here we take $b=1$). The effect of $T$ is even more interesting. For high values of $T$, defectors out-compete  cooperators, spread in the population and, as a consequence, the social wealth decreases. Thus, the overall birth rate is lowered and the population density decreases. The dependence of the population density on $T$ and $d$ is shown in details in figure  \ref{fig2}, where results from simulations are compared to the mean-field approximation.  Notice that, in a homogeneous population of defectors, the payoff of all individuals are all equal to zero and the population density is no longer affected by the values $T$, as shown in Fig \ref{fig2}-b.

\begin{figure}
\centering
\includegraphics[scale=0.6]{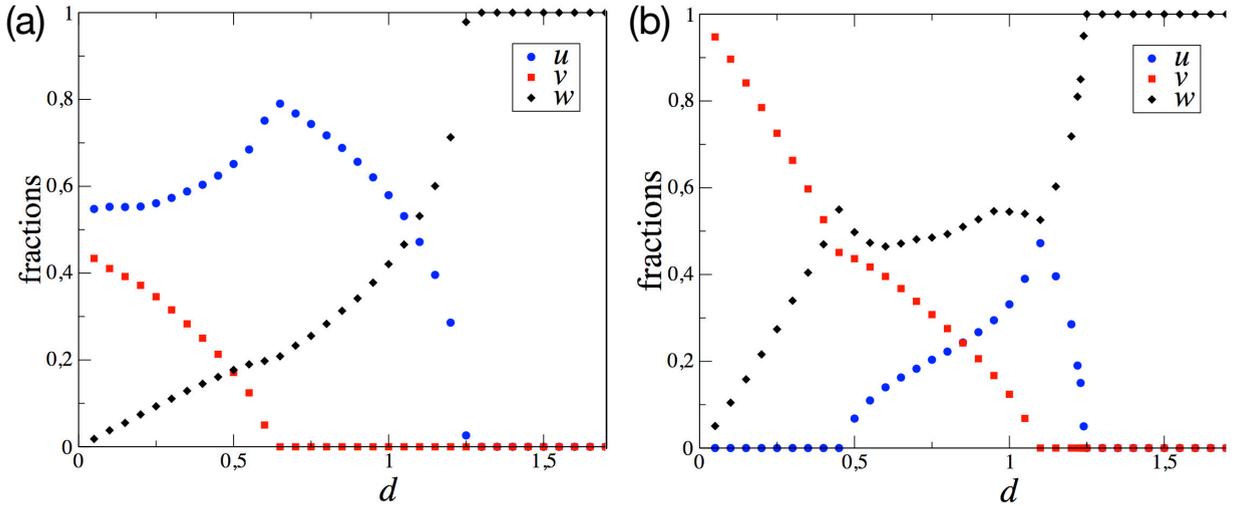}
\caption{Composition of the population in the stationary state. The figure shows the fraction of cooperators (circle, blue), defectors (square, red), and vacancies (square, black)  in the square lattice. Panel (a) is for $T=2$ and panel (b) is for $T=4$. Notice that the fraction of defectors decreases monotonically, whereas the fraction of cooperators exhibits non-monotonic behavior. In particular, if the defection tendency $T$ is sufficiently high, cooperators can survive only for moderate values of the death rate. In all cases, the population is extinct if the death rate is sufficiently high. }
\label{fig3}
\end{figure}

The fraction of cooperators, defectors and empty sites in the stationary state are shown in Fig.  \ref{fig3}. If $T=1$,  defectors are less competitive because they do not produce any payoff and cannot exploit cooperators. Thus, only cooperators  survive in the stationary state.  As expected, the density of cooperators decreases as  $d$ increases (not shown in the figure). For $T>1$, defectors and cooperators can coexist.  The fraction of defectors decreases as $d$ increases and, for sufficiently high $d$ values, defectors are extinct.    
More interestingly, the fraction of cooperators exhibits non-monotonic behavior. The fraction of cooperators increases as $d$ moves away from zero, reaches a peak, and decreases as $d$ gets higher. Notice that the fraction of cooperators reaches its peak at a moderate value of $d$. If the temptation to defect is high, this non-monotonic behaviour is more remarkable: cooperators are extinct at low $d$ values, survive for intermediate $d$ values, and are extinct for high $d$ values. In general, moderate death rates are beneficial for cooperation. We can understand this non-monotonic behavior  phenomenon as follows. If the death rate is low,  the population is less fragmented and defectors can exploit cooperators. Consequently, cooperators do worse than defectors. As $d$ gets higher, the fragmentation increases. Since offspring remains close to the parents, the formation of cooperator's clusters starts to provide an advantage to the cooperators, boosting the overall level of cooperation. Consequently, the population density stops decreasing and remains relatively stable at a higher level, as shown in Fig. \ref{fig1}. Recall that cooperators increase the population average payoff to levels that are higher than that in a homogeneous population of defectors.  Surely, if $d$ is sufficiently high, both cooperators and defectors are extinct. Interestingly,  defectors are extinct at $d$ values that are lower than the $d$ values at which cooperators are extinct.  We note that, in competitive scenarios, there is a general phenomenon where harsh external conditions create an evolutionary advantage to cooperators. This effect is very broad, and have been produced with different mechanisms \cite{death1,death2,alonso2006}. An interesting example can be seen in  the stalk construction of \textit{Dictyostelium discoideum}, a species of phagotrophic bacterivores \cite{travisano2004}. The response to starvation crucially depends on cooperative behavior, where some cells form a stalk so the spores have the chance to survive somewhere else. If the starvation is severe, which increases the death probability, cooperators must be present if the population is to have a chance of survival.  In other words, when death knocks the door,  defectors are the first to perish, whereas cooperators stand longer due to  mutual support.

The interaction between cooperators and defectors give rise to oscillatory dynamics in finite populations at high values of $T$. To characterize oscillatory behavior, we performed a Fourier analysis. The Fourier spectrum of the time series $f(t)=u(t)+v(t)$  is given by
\begin{equation}
A_f(\omega)=\lim_{T\rightarrow\infty}\frac{1}{T}\left|\sum^T_{t=1}f(t)e^{i\omega t}\right|^2~~.
\end{equation}
The presence of oscillations is reflected as a peak at a nonzero frequency in the Fourier spectrum. A non-homogeneous distribution in the amplitudes (higher amplitudes for low frequencies) distinguish oscillatory regions from non-oscillatory ones \cite{antal2001}. In non-oscillatory regions, the amplitudes are equally distributed and we expect the average value of $A$ to be low. In contrast, in oscillatory regions, the values of $A$ will concentrate at low frequencies and the average value of $A$ is expected to be higher. Hence, to characterize oscillatory regions we calculated averages over the amplitudes related to non zero frequencies and we observed that $\bar A>0.02$ is a good indicator of oscillations.

Figure 4 shows a typical Monte Carlo run in the oscillatory region. First, the fraction of cooperators increases, creating a wealthy environment. The defectors exploit the cooperators, degrading the environment. The net effect is that there is a reduction in the birth rate, increasing the number of vacancies. As a consequence,  cooperators get isolated in islands,  depriving the defectors from  resources, and the cooperators start to grow again.  We note that, in the oscillatory region, the cycle between the three states (cooperator, defector and vacancy) is  coupled: the peak of one state coincides with the valley of the other. The vertical dashed lines in Fig. 4 show some of the many cycles, where the peak in cooperation coincides with the minimum of vacancies. This effect can be seen as cyclic dominance between three species, like in the paper-rock-scissor game \cite{cannon2007}. If we consider vacancies as another kind of strategy, we can understand these cycles as a process where defectors take advantage of the cooperators, leading to an environment with a low payoff. This, in turn, enables vacancies to grow and defectors start to die. In this new environment full of vacancies, cooperators are again free  to grow because they can produce payoff due to mutual cooperation (recall that defectors do not produce any payoff). It is also important to notice that all three oscillations are highly coupled, following a precise cyclic order:  defectors $\rightarrow$ vacancies  $\rightarrow$ cooperators $\rightarrow$ defectors. We also note that, since the population is finite, oscillations may drive the population to extinction.

\begin{figure}
\centering
\includegraphics[scale=0.6]{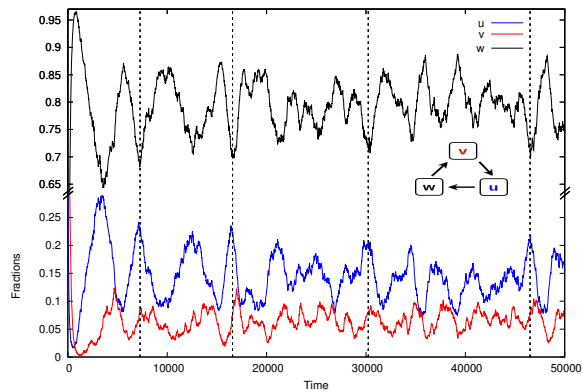}
\caption{Time evolution of the fractions of cooperators and defectors for $T=10$ and $d=1.1$. The oscillation of defectors (red) is late compared to the oscillation of cooperators (blue). Since the population is finite, large fluctuation may drive the population to extinction (not shown in the figure). The dashed lines indicate that the peak of cooperation coincides with a maximum of the population density. }
 \label{osc}
\end{figure}

This kind of cyclic dominance has been studied in rock-papers-scissor games and, more recently,  in multi-strategy situations \cite{Perc2015, Szolnoki2017}. Here, this behaviour is an emergent one and happens only within a very specific window of parameters. It is also important to notice that this microscopic mechanism can explain the unusual peaks observed in Figure 3, where for some parameters, there are primary and secondary peaks of vacancies and cooperators. More specifically, the peak of cooperation  happens when its predator (defectors) are extinguished. But cooperators cannot maintain this peak because, without defectors, vacancies start to grow unchecked. This process is very similar to what is observed in three species dynamics, where the absence of a predator can  lower its prey population if there is a third species involved \cite{lutz_jtb13, rulquin_pre14}. It is also important to note that, in cyclic games, the oscillatory behaviour  is a feature of systems with finite size \cite{ciclic}. In biological models, the systems are finite and the oscillatory behavior is expected to play an important role in the population dynamics. We analysed the dependence of the amplitude of oscillation on the system size, more specifically, we measured the dependence of the variance of the time series on the system size. As expected, the amplitudes decrease as the size increases, indicating that the cyclic behaviour seen in our model is an emergent feature in systems of finite size.

\begin{figure}
\centering
\includegraphics[scale=0.5]{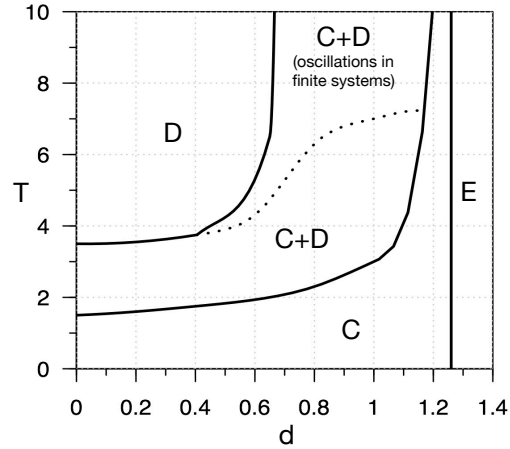}
\caption{Phase diagram in the stationary state for the whole $T-d$ plane. The parameter space is  divided into regions where only cooperators survive (C), only defectors survive (D), cooperators and defectors coexist (C+D), or the population is extinct (E). The dotted line delimits the region of coexistence where we found oscillatory behaviour in finite-size systems. Since the oscillatory behaviour vanishes in the limit of infinite population size, the dotted line does not represent any kind of phase transition.}
\label{parameter}
\end{figure}

To sum up, we show in Fig. \ref{parameter} the stationary state phase diagram. The stationary state of the population is classified into four phases in the parameter space $T \times d$: pure cooperators; pure defectors;  coexistence of cooperators and defectors; and empty population. The dependence of cooperation on the death rate along a horizontal line defined by $T=5$ in Fig. \ref{parameter} illustrates  the transitions:  for low death rates, cooperators have no chance against defectors; for higher death rates, cooperators rise and coexist with defectors; for even higher death rates, only cooperators survive; finally, as expected, if the death rate is sufficiently high, the population is extinct. Interestingly, if the temptation to defect is much higher ($T=10$, for instance) and the death rate is moderate, the coexistence between cooperators and defectors is characterized by an oscillatory behaviour in finite systems. However, the amplitude of oscillations decreases with system size and, in the limit of infinite population, the time series becomes smoother and converge to stationary values. We should stress that analysis of finite systems is relevant for biological applications, since real populations are always finite.

\section{Conclusions}
\label{conclusions}

The compromise between the simplicity of the model and the biological reality must always be calibrated so important phenomena are not neglected.  The modelling of the evolution of cooperation with fluctuating population size exhibits interesting results. The evolutionary outcome is very rich, ranging from homogeneous populations to heterogeneous populations where cooperators and defectors co-exist. In finite systems, there can be oscillatory behavior.

The population density is trivially affected by the  death rate if the temptation to defect is small: the higher the death rate, the lower the population density. For higher temptation values, where cooperation does not survive at how values of the death rate, the population density interrupts the decrease for death rate values near the cooperation resurrection point. The effect of the temptation to defect is more subtle. An increase in the temptation to defect favours defectors over cooperators. Since a population with more defectors has a lower overall birth rate, the population density decreases. 

In spatial populations with limited dispersal of the offspring, clusters of cooperators provide support to cooperators in the competition against defectors. If the temptation to defect is sufficiently high, the defectors can invade the clusters of cooperators, exploit the cooperators and overcome the cooperators. Such an exploitation mechanism is efficient as long as the clusters are stable.  However, if the death rate is moderate, the clusters are constantly fragmented and individuals often become more isolated. The growth of new clusters is more beneficial to cooperators because the cluster of defectors do not produce any payoff. Hence, moderate death rates can be beneficial for cooperators.

\appendix
\section{Calculation of the fixed points in the mean field approximation}

The first fixed point is the trivial solution, where ${u=v=0}$ and ${w=1}$.

The second fixed point satisfies  $u=0$, $v\neq0$, and $w\left(b+U_D\right)-d=0$.
So we have
\begin{eqnarray*}
 w&=&\frac{d}{b+U_D}=\frac{d}{b}\\
u&=&0\\
 v&=&1-w=1-\frac{d}{b}.
\end{eqnarray*}

The third fixed point satisfies $v=0$, $u\neq0$, and $w\left(b+U_C\right)-d=0$. So we have
\begin{eqnarray*}
 w&=&\frac{b+4-\sqrt{\left(b+4\right)^2-4d}}{2}\\
 u&=&1-w \\
 v&=&0 
\end{eqnarray*}

The fourth fixed point satisfies $u\neq0$, $v\neq0$ and
\begin{eqnarray*}
 w\left(b+U_C\right)-d &=& 0 \\
 w\left(b+U_D\right)-d &=& 0.
\end{eqnarray*}
So we have
\begin{eqnarray*}
 \frac{d}{U_C+b} &=& \frac{d}{U_D+b} \nonumber \\
 U_C &=& U_D \nonumber \\
 u &=& Tu \nonumber~~~~.
\end{eqnarray*}
This requirement is satisfied  if $u=0$, which is not acceptable since we are assuming that $u\neq0$. So one must have $T=1$, yielding
\begin{eqnarray*}
u&=&u^*\\
w&=&\frac{d}{b+4u^*} \\
v&=&1-\frac{d}{b+4u^*}-u^*
\end{eqnarray*}


\end{document}